\documentstyle[12pt,epsf]{article} 
\normalbaselines
\begin{document}
\bibliographystyle{unsrt}

\begin{center}
{\tiny ``{\it Fifth International Conference on Squeezed States and
Uncertainty Relations}", Balatonfured, Hungary (May 1997).\\ D.~Han,
J.~Janszky, Y.~S.~Kim and V.I.~Man'ko (Editors),\\ NASA/CP-1998-206855,
pages 583-588.}
\end{center}

\vbox {\vspace{3mm}} 

\begin{center}
{\large \bf DYNAMICAL MANIPULATION FOR SPIN-1/2 SYSTEMS}\\[7mm]
David J.~Fern\'andez C. and Oscar Rosas-Ortiz\\
{\it Departamento de F\'\i sica, CINVESTAV-IPN \\
A.P. 14-740, 07000 M\'exico D.F., MEXICO}
\end{center}

\vspace{2mm}  

\begin{abstract}
By means of the inverse techniques we analyse the evolution of purely
spin-$1/2$ systems in homogeneous magnetic fields as well as the
generation of exact solutions. Some {\it evolution loops}, dynamical
processes for which any state evolves cyclically, are presented, and their
corresponding geometric phases are evaluated.
\end{abstract}

In the direct approach to quantum dynamics, the main goal is to look for
solutions $\vert\psi(t)\rangle \in S({\cal H})$ to Schr\"odinger's
equation
\begin{equation}
i\hbar \frac{d}{dt}\vert\psi(t)\rangle = H(t) \vert\psi(t)\rangle,
\end{equation}
provided the Hamiltonian $H(t)$ and the initial state $\vert\psi
(0)\rangle$ are given, where $S({\cal H}) = \{\vert \psi\rangle\in{\cal H}
\ \vert \ \langle \psi\vert \psi\rangle = 1\}$, ${\cal H}$ is the system
Hilbert space [1]. The same ideology is present in the study of the
`shadow' motion of $\vert\psi(t)\rangle$ on projective Hilbert space
${\cal P}$, where $p\in{\cal P}$ represents a possible physical state.
${\cal P}$ is obtained through a projective map $\pi :  S({\cal
H})\rightarrow {\cal P}$, and all $\vert\psi'\rangle =
e^{i\phi}\vert\psi\rangle, \ \phi\in{\bf R}$ (a ray of $S({\cal H})$), are
carried into the same $p\equiv\pi(\vert\psi\rangle) \in {\cal P}$. In this
way, $S({\cal H})$ is seen as a fibre bundle with base space ${\cal P}$
and fibre $U(1)$ [2-3].

For spin-$1/2$ systems in magnetic fields ${\bf B}(t)$, the spin part of
the Hamiltonian reads: 
\begin{equation}
H(t) = -\mu{\bf B}(t)\cdot{\bf S} = - \frac{\mu\hbar}{2}{\bf B}(t)\cdot
\mbox{\boldmath $\sigma$} ,
\end{equation}
where ${\bf S}$ is the spin operator, $\mbox{\boldmath
$\sigma$}$-components are Pauli matrices, $S({\cal H})$ is a sphere $S^3$
embeded in ${\bf R}^4$, ${\cal P}$ is a sphere $S^2$ embeded in ${\bf
R}^3$, and $\pi(\vert\psi\rangle)={\bf n} \equiv \langle\psi\vert
\mbox{\boldmath $\sigma$}\vert\psi\rangle \in S^2$. The evolution equation
for ${\bf n}(t)$ is obtained from (1-2): 
\begin{equation}
\frac{d{\bf n}(t)}{dt} = -\mu {\bf B}(t) \times {\bf n}(t).
\end{equation} 
Equations (1-3) are usually solved by means of numerical integration. 

It is interesting also to look for exactly solvable models where the
relevant information would be found in one analytic expression. However,
it is hard to find them due to the complexity to sum up the `continuous'
Baker-Campbell-Haussdorf exponent arising when the evolution operator is a
true exponential [4]. It is well known that exact solutions to (1-3)
arise when: {\it i)} the field has a fixed direction, ${\bf B}(t) =
B(t){\bf e}_B$;  {\it ii)} ${\bf B}(t)$ rotates around a fixed direction
with constant angular velocity, and $\vert{\bf B}(t)\vert = \vert{\bf
B}(0)\vert$. In the first case, by solving the eigenproblem for the
constant matrix $\sigma_B = \mbox{\boldmath $\sigma$}\cdot{\bf e}_B$ and
expanding then $\vert\psi(0)\rangle$ in that basis the general solution
$\vert\psi(t)\rangle$ is found.  For the rotating field, the key is the
`transition to the rotating frame' where it arises a constant effective
Hamiltonian, which can be solved as previously [5]; by coming back to
the initial frame the solution is gotten. 

For general ${\bf B}(t)$ one can try again the transition to the rotating
frame in which the direction of ${\bf B}(t)$ would be fixed though the
effective Hamiltonian won't be constant; thus, this problem generally
cannot be solved (see, e.g., [6]). This is the motivation to look for
alternative techniques to circumvent these difficulties.

In this work, a method to generate solutions for (1-3) is studied.  We
will use inverse techniques, i.e., we are going to look for the magnetic
fields inducing a given ${\bf n}(t)$. This subject, introduced by Lamb
[7], has been pursued with different names (inverse problem, control
theory, dinamical manipulation, etc. [8-11]). We use {\it dynamical
manipulation} because somehow it has implicit our ideals:  to enforce
systems to evolve as we want. 

Suppose that a spin state ${\bf n}(t)\in S^2$ is given. The system of
equations (3) for ${\bf b}(t)\equiv\mu{\bf B}(t)$, $M({\bf n}){\bf b}(t) =
\dot{\bf n}(t)$, is such that ${\rm det}M({\bf n})=0$. Thus, one ${\bf
n}(t)$ does not lead to a unique ${\bf b}(t)$. If $b_3(t)$ is left
arbitrary the other components are: 
\begin{equation}
b_1(t) = [b_3(t)n_1(t) + \dot{n}_2(t)]/ n_3(t) , \quad
b_2(t) = [b_3(t)n_2(t) - \dot{n}_1(t)]/ n_3(t).
\end{equation}
As ${\bf n}(t)=R(t){\bf n}(0)$, ${\bf b}(t)$ depends of the generic motion
(encoded in $R(t)\in SO(3)$) and the initial condition. This is
unsatisfactory because two different fields would be needed to induce two
trajectories with common $R(t)$ but different ${\bf n}(0)$. However, in
the direct approach a constant field leads unambiguously to $R(t)$. Would
it be possible that one $R(t)$ determines a unique ${\bf b}(t)$ when the
inverse techniques are applied? This is analysed below for specific forms
of $R(t)$.

Let us suppose first that ${\bf n}(t)$ rotates with angular velocity
$\dot{\delta}(t)$ around $z$-axis:  
\begin{equation}
{\bf n}(t) = \sin\theta_0\cos(\phi_0+\delta(t)){\bf i} +
\sin\theta_0\sin(\phi_0+\delta(t)){\bf j} + \cos\theta_0{\bf k},
\end{equation}
where ${\bf i}, \ {\bf j}, \ {\bf k}$ are unit vectors along $x, \ y, \
z,$ $\delta(0)=0$. The fields inducing (5) become: 
\begin{equation}
{\bf b}(t) =\tan \theta_0 \,[ b_3(t) + \dot{\delta} (t)]  
[\cos ( \phi_0 + \delta (t) ) {\bf i} + \sin ( \phi_0 + \delta (t) )  
{\bf j}] + b_3(t) {\bf k}.  
\end{equation}
It is clear that ${\bf b}(t)$ depends on ${\bf n}(0)$; to avoid this we
make $b_3(t) = - \dot{\delta}(t)$. Hence: 
\begin{equation}
{\bf b}(t) = - \dot{\delta}(t){\bf k}.
\end{equation}
If $\dot{\delta}(t) = \omega$, we recover the standard solution for the
spin in a constant field.

Let us analize a more general case including the standard magnetic
resonance model [5]. To this aim, suppose that ${\bf n}(t)$ rotates
simultaneously around two fixed directions with variable angular
velocities. Let us choose one of these directions along ${\bf k}$ and the
other one along a vector ${\bf e}_\chi$ on $x-z$ plane at an angle $\chi$
from ${\bf k}$. The rotation matrix is: 
\begin{equation}
R(t) = R_3(\beta(t))R_\chi(\alpha(t)) = R_3(\beta(t)) R_2^{-1}(-\chi) 
R_3(-\alpha(t))R_2(-\chi) ,
\end{equation}
where $R_2(\omega)$ and $R_3(\omega)$ are finite rotation by $\omega$
around ${\bf j}$ and ${\bf k}$. If the $R_2(\omega)$ and $R_3(\omega)$ of
[1] are used to evaluate ${\bf n}(t)= R(t) {\bf n}(0)$ and this is
substituted into (4) one arrives at:
\begin{eqnarray} 
b_1(t) &=& [-\sqrt{1-\lambda^2}N_2(t) + \lambda N_3]^{-1}\Bigl\{ \sin
\beta(t) \, \{\lambda\mathaccent95 \alpha(t) -
\bigl[b_3(t) + \mathaccent95 \beta (t)\bigr]\} \, N_1(t) \nonumber \\
&+& \cos \beta(t) \, \bigl\{ \bigl[b_3(t) +\mathaccent95 \beta (t)\bigr]
[\lambda N_2(t) + \sqrt{1-\lambda^2}N_3] - \dot\alpha(t)N_2(t)\}\Bigr\}, 
\end{eqnarray}
\begin{eqnarray} 
b_2(t) &=& [-\sqrt{1-\lambda^2}N_2(t) + \lambda N_3]^{-1}\Bigl\{ -\cos
\beta(t) \, \{\lambda\mathaccent95 \alpha(t) -
\bigl[b_3(t) + \mathaccent95 \beta (t)\bigr]\} \, N_1(t) \nonumber \\
&+& \sin \beta(t) \, \bigl\{ \bigl[b_3(t) +\mathaccent95 \beta (t)\bigr]
[\lambda N_2(t) + \sqrt{1-\lambda^2}N_3] - \dot\alpha(t)N_2(t)\}\Bigr\}, 
\end{eqnarray}
where $\lambda\equiv\cos\chi$, ${\bf n}(0) \equiv (x_0,y_0,z_0)$, and:
\begin{equation}
\begin{array}{l}
N_1(t) = -\lambda \sin \alpha(t) x_0 + \cos \alpha (t) y_0 +
\sqrt{1-\lambda^2}\sin \alpha(t) z_0,
\nonumber\\
N_2(t) = \lambda \cos \alpha (t) x_0 + \sin \alpha(t) y_0 - 
\sqrt{1-\lambda^2}\cos \alpha(t) z_0,
\nonumber\\
N_3  = \sqrt{1-\lambda^2} x_0 + \lambda z_0 .
\nonumber
\end{array}
\end{equation}
The field independent of ${\bf n}(0)$ appears if $b_3(t) + \mathaccent95
\beta(t) = \lambda\mathaccent95 \alpha(t)$:
\begin{equation}
{\bf b}(t) = \dot{\alpha}(t) \sqrt{1-\lambda^2} \,\bigl[\cos \beta(t){\bf i} + 
\sin \beta(t){\bf j}\bigr] +  \bigl[\lambda \dot{\alpha}(t) - 
\dot{\beta}(t)\bigr]{\bf k}. 
\end{equation}
There is a consistency condition, ${\bf n}(0) = R(0) {\bf n}(0) \
\Rightarrow$ $R(0) = 1 \ \Rightarrow \alpha (0) = \beta(0) = 0$. 

We pass now to the particular cases, firstly those of the introduction. On
the one hand, our solution (8-12) provides in two ways the case when
${\bf n}(t)$ rotates around ${\bf k}$: by taking $\chi = 0$, $\lambda =
1$, and ${\bf b}(t)  = \bigl[ {\dot\alpha}(t)-{\dot \beta}(t)\bigr]{\bf
k}$; take now $\alpha(t) = 0$, which leads to ${\bf b}(t) =
-{\dot\beta}(t){\bf k}$. On the other hand, if the rotations of ${\bf
n}(t)$ are uniform ($\alpha(t) = \alpha_0 t$, $\beta(t) = \beta_0 t$),
we arrive at the traditional model used to examine the magnetic resonance
[5]:
\begin{equation}
{\bf b}(t) = \alpha_0\sqrt{1-\lambda^2}\,\bigl[\cos(\beta_0t){\bf i} + 
\sin(\beta_0t){\bf j}\bigr] + (\lambda\alpha_0 -\beta_0){\bf k}.
\end{equation}
The resonance explanation runs as follows. Suppose that the spin points
along ${\bf k}$, and it is placed in a constant field ${\bf B}_0=B_z{\bf
k} = (\lambda\alpha_0 - \beta_0){\bf k}/\mu$. Hence, at $t=0$ the spin
state is an eigenstate of the `base' Hamiltonian $H_0 = -(\lambda\alpha_0
- \beta_0)S_z$, $\vert\psi(0)\rangle = \vert +\rangle$, where $S_z\vert +
\rangle = (\hbar/2)\vert + \rangle$. On $S^2$ we have ${\bf n}(0) =
(0,0,1)$. Now, at $t=0$ we superimpose to ${\bf B}_0$ the field ${\bf
B}_\perp = (\alpha_0/\mu) \sqrt{1-\lambda^2} [\cos(\beta_0 t){\bf i} +
\sin(\beta_0t){\bf j}]$. Thus, a formally `perturbative Hamiltonian' $W =
-\alpha_0\sqrt{1 - \lambda^2}[ \cos(\beta_0 t)S_x +\sin(\beta_0t)S_y]$ is
gotten (permitting however the exact treatment).  It could induce at $t>0$
a transition to the orthogonal eigenstate $\vert - \rangle$ of $H_0$,
which on $S^2$ is represented by the vector ${\bf n}_- = (0,0,-1)$. The
probability transition is: 
\begin{equation}
P_{+\rightarrow -}(t) \equiv \vert \langle - \vert \psi(t)\rangle\vert^2 = 
[1-n_3(t)]/2 = (1-\lambda^2)\left[1-\cos(\alpha_0 t)\right]/2.
\end{equation}
Notice that $P_{+\rightarrow -}(t)$ is small if $\chi$ is small or
$\chi\approx \pi$, i.e., when the rotations axes of ${\bf n}(t)$ are
aligned. The greatest value arises for $\chi = \pi/2$ (orthogonal rotation
axes): 
\begin{equation}
P_{+\rightarrow -}(t) = \left[1 - \cos(\alpha_0 t)\right]/2.
\end{equation}
At $\tau_n = (2n + 1)\pi/\alpha_0$ this probability is $1$, i.e., the
state certainly will be $\vert - \rangle$. This resonance is understood
analysing the physics in the rotating frame: the rotating observer will
see that (due to a non-inertial field produced by the rotation) the level
spacing between $\vert + \rangle$ and $\vert - \rangle$ will decrease with
respect to the inertial observer. For $\lambda = 0$ this spacing will be
zero because the non-inertial field cancels ${\bf B}_0$, and the
transition in the eyes of the rotating observer certainly will be induced.
In terms of $S^2$ there is an equivalent explanation: in the rotating
frame it is present a constant effective field ${\bf b}_{eff} =
\alpha_0(\sqrt{1-\lambda^2}\,{\bf i}+\lambda{\bf k})$ around which ${\bf
n}(0)= (0,0,1)$ starts to precess. But if ${\bf n}(0)$ has to be converted
into ${\bf n}_- = (0,0,-1)$, ${\bf b}_{eff}$ should have vanishing $z$
component ($\lambda=0$). In this case, ${\bf n}(t)$ will simply precess
around ${\bf i}$ with angular velocity $\alpha_0$ (period
$T=2\pi/\alpha_0$), and at $\tau_n = T/2 + nT=(2n+1)\pi/\alpha_0$ the spin
state will be ${\bf n}(\tau_n) = n_- = (0,0,-1)$.

Let us remind the absence of essential restrictions on the real functions
$\alpha (t)$ and $\beta(t)$ in (8-12). This makes possible to study a lot
of other interesting examples, e.g., the case when the two rotations
(8) would be induced by a field with a constant third component,
$b_3(t) = b_0$, which leads to $\alpha(t) = [b_0 t + \beta(t)]/\lambda$. 
Hence:
\begin{equation}
{\bf b}(t) = \frac{\sqrt{1-\lambda^2}}{\lambda}\bigl[b_0 + {\dot\beta}(t) 
\bigr]\bigl[\cos\beta(t){\bf i} + \sin\beta(t){\bf j}\bigr] + b_0{\bf k}. 
\end{equation}
This field has a $x-y$ projection rotating around ${\bf k}$ with angular
velocity ${\dot\beta}(t)$ and amplitude $\vert b_0 + {\dot\beta}(t) 
\vert\sqrt{1/\lambda^2-1}$. It represents a new resonance model analogous
to (13-15) where the zero Hamiltonian is associated to $b_0{\bf k}$ and
the `perturbation' corresponds to the field component orthogonal to ${\bf
k}$. Let us suppose that ${\bf n}(0)  = (0,0,1)$ and evaluate the
probability that at time $t$ the system will be in ${\bf n}_-=(0,0,-1)$: 
\begin{equation}
P_{+\rightarrow -}(t) = [1-n_3(t)]/2 = (1-\lambda^2)
 \left[1-\cos\alpha(t)\right]/2.
\end{equation}
Once again, the greatest probability arises for $\lambda = 0$: 
\begin{equation} 
P_{+\rightarrow -}(t) = [1-\cos\alpha(t)]/2,
\end{equation} 
and at $\tau_n$ such that $\alpha(\tau_n) = (2n+1)\pi$, $P_{+\rightarrow
-}(\tau_n) = 1$, i.e., the system certainly will make the transition from
$\vert +\rangle$ to $\vert -\rangle$ due to the `perturbation'
$W=-[b_1(t)S_1 + b_2(t) S_2]$. 

It is interesting to study the possibility that any ${\bf n}(t)\in S^2$
would be cyclic, i.e., that the evolution operator $U(t)$
($\vert\psi(t)\rangle = U(t) \vert\psi (0)\rangle$) would be the identity
(modulo phase) at some $t=\tau$. These processes, named {\it evolution
loops} (EL) [8,11-12], mimic the harmonic oscillator behaviour for
systems with time-dependent Hamiltonians. The EL are useful to manipulate
quantum systems, e.g., to induce the squeezing of the wavepacket inside of
a Penning trap variant [11].  They are helpful as well to produce the
rigid displacement of the wavepacket inside a `magnetic chamber' perturbed
by homogeneous time-dependent electric fields [11]. Some arguments
indicate that, by applying perturbations, they can be used to induce any
unitary operator as the result of the precession of the distorted loop
[8]. 

For $1/2$ spin systems, the EL provide non-trivial cyclic evolutions for
which the geometric phase can be explicitly evaluated [13] (see also [12]
and below). In our example (8-12), the loop condition is immediately
formulated:  the EL exist if there is a time $\tau$ for which ${\bf n}(t)$
simultaneously performs $n$ effective rotations around ${\bf k}$ and $l$
around ${\bf e}_\chi$, $\alpha(\tau) = 2l\pi , \ \beta(\tau) = 2n\pi, \
\tau > 0$, $n\in{\bf Z}, \ l\in{\bf Z}$. This condition translates into
some equations for the field-spin parameters and $\tau$. It was proposed
in [13] for the rotating field (13), and an example of a hypocycloid
performing $1$ rotation around ${\bf k}$ and $5$ rotations around ${\bf
e}_\chi$ was shown\footnote{There it was explored also the EL for the spin
in the oscillating field ${\bf b}(t) = [b_0 + b\cos(\omega t)]{\bf k}.$}. 
Other authors obtained interesting results for the same physical system
[14]. Here, we want to show `deformed versions' of the figure in [13], for
a field of form (16) with $\beta(t) = a\lambda\sin(\alpha_0 t) +
(\lambda\alpha_1 - b_0)t$. Our results are drawn in figure 1 for two
essentially different EL with $\tau=2\pi$, $b_0 = 3, \ \lambda = 4/5$: 
{\it a)} first we choose $a = 0.7, \ \alpha_1 = \alpha_0 = 5$; {\it b)}
now we choose $a = 1, \ \alpha_1 = 5 - 1/2\pi, \ \alpha_0=9/4$ .

\centerline{\epsfbox{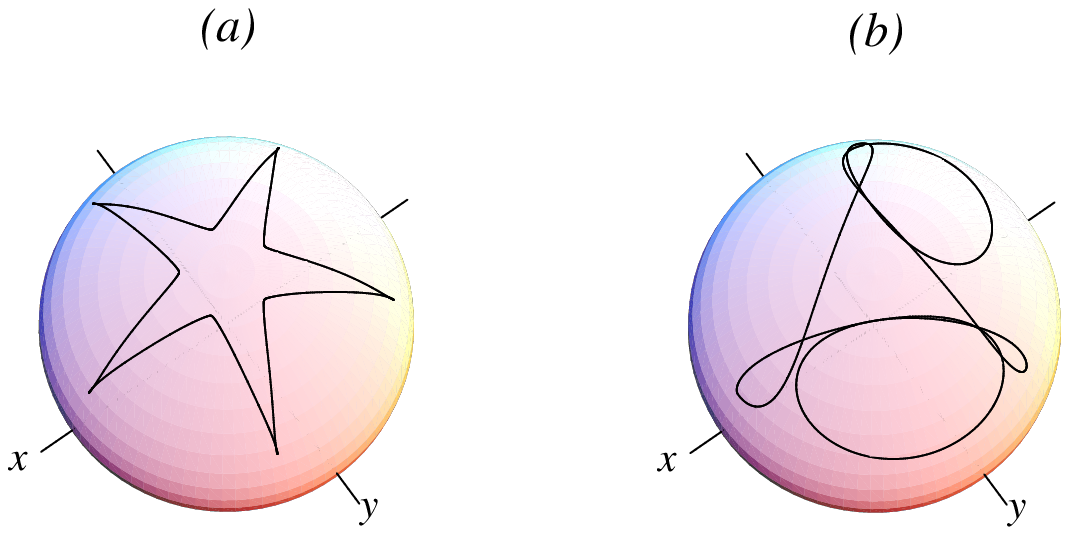}}
\begin{quotation}
Fig.1. Spin trajectories illustrating two EL of period $\tau=2\pi$
induced
by (16) with $\beta(t) = a\lambda \sin(\alpha_0 t) + (\lambda\alpha_1 -
b_0)t$, $b_0 = 3, \ \lambda = 4/5, \ {\bf n}(0) = (\sqrt{3}/2 , 0, 1/2)$,
and: {\it a)} $a = 0.7, \ \alpha_1 = \alpha_0 = 5$;  {\it b)} $a = 1, \
\alpha_1 = 5 - 1/2\pi, \ \alpha_0 = 9/4$. 
\end{quotation} 

It is easy to see that case {\it a} provides a periodic loop: 
$U(\tau=2\pi)=1$ and $U(m\tau)=1, \ m\in{\bf N}$. This happens because the
Hamiltonian, for this choice of parameters, is periodic with period
$\tau=2\pi$. For case {\it b} the parameters are such that
$U(\tau=2\pi)=1$, but this EL is not periodic, $U(2\tau)\neq 1$, because
the Hamiltonian is not. Different examples of this kind of aperiodic loop,
when the field oscillates along ${\bf k}$, are given elsewhere [13].
 
Now, if an EL is produced at $t=\tau$, then any state ${\bf n}(t)$ becomes
cyclic, ${\bf n}(\tau)= {\bf n}(0)$. Thus, it makes sense to evaluate the
corresponding geometric phases, which characterize some global curvature
effects of ${\cal P}=S^2$ [3,12-15]. For spin-$1/2$ systems the
geometric phase turns out to be minus one half of the solid angle
$\Delta\Omega$ subtended by the oriented closed curve ${\bf n}(t)\in S^2$
[13], where: 
\begin{equation} 
\Delta\Omega = \int_0^\tau\frac{n_1{\dot n}_2 -n_2{\dot n}_1}{1+n_3}dt. 
\end{equation} 
For our case (8-12), this formula can be evaluated using the loop
condition:
\begin{equation}
\Delta\Omega = 2n\pi[1 - \cos\chi\cos(\theta-\chi)] -
2l\pi[1 - \cos(\theta - \chi)] + \sin\chi\sin(\theta -
\chi)\int_0^\tau{\dot\beta}\cos\alpha dt,
\end{equation}
where we have taken ${\bf n}(0)=(\sin\theta,0,\cos\theta)$.  This
expression is equal to the corresponding one for the rotating field (13)
but for the last term [13], which depends on the explicit form of both
$\alpha(t)$ and $\beta(t)$. If $b_3(t)=b_0$ (the case discussed at
(16-18)) we get:
\begin{equation}
\Delta\Omega = 2n\pi[1 - \cos\chi\cos(\theta-\chi)] - 2l\pi[1 - 
\cos(\theta - \chi)] -b_0 \sin\chi\sin(\theta - \chi)\int_0^\tau\cos\alpha
dt,
\end{equation}
and the integral depends just of $\alpha(t)$. Notice that this integral
vanishes if $\alpha(t) = \alpha_0 t$ and $\alpha_0\tau = 2l\pi$, which
leads once again to the known result for the EL in the standard case (13). 
The simplest situation leading to EL for which the integral in (21) does
not vanish arises when $\alpha(t)$ is quadratic in $t$. With this choice,
that term will involve Fresnel integrals\footnote{Take $\alpha(t) =
\alpha_0 t^2$ with $\alpha_0 = 5(2\pi)^{-1}$, $b_0=3, \ \lambda=4/5$, to
produce an EL such that $\alpha(\tau=2\pi) = 10\pi, \ \beta(2\pi) = 2\pi$.
Moreover, $\int_0^{\tau=2\pi}\cos\alpha(t) dt \equiv
\pi(5)^{-1/2}C(2\sqrt{5})= 0.700896 \neq 0$.}. 

In conclusion, the manipulation techniques are appropriate to generate
solvable examples for the spin evolution in homogeneous magnetic fields.
When these techniques are applied to produce the EL, it arises cyclic
evolutions for which the geometric phases can be explicitly evaluated. We
hope that our treatment has shed as well some light on the functioning of
the resonance mechanism.

\smallskip

The authors acknowledge the support of CONACYT (M\'exico).

\end{document}